\documentclass[entropy,article,submit,moreauthors,pdftex,12pt,a4paper]{mdpi} %
\lastpage{x}
\doinum{10.3390/------}
\pubvolume{xx}
\pubyear{2013}
\history{Received: xx / Accepted: xx / Published: xx}

\usepackage{amssymb,amsmath}
\usepackage{graphicx}
\Title{Enhanced sampling in molecular dynamics using metadynamics,
replica-exchange, and temperature-acceleration}

\Author{Cameron Abrams $^{1,\star}$ and Giovanni Bussi $^{2}$}

\address{%
$^{1}$ Department of Chemical and Biological Engineering, Drexel University, 3141 Chestnut St., Philadelphia, Pennsylvania, United States  19104\\
$^{2}$ Scuola Internazionale Superiore di Studi Avanzati (SISSA), via Bonomea 265, 34136 Trieste, Italy}

\corres{cfa22@drexel.edu; tel +1-215-895-2231}

\abstract{ 
We review a selection of methods for performing enhanced
  sampling in molecular dynamics simulations.  We consider methods
  based on collective variable biasing and on tempering, and offer
  both historical and contemporary perspectives.  In
  collective-variable biasing, we first discuss methods stemming from
  thermodynamic integration that use mean force biasing, including the
  adaptive biasing force algorithm and temperature acceleration.  We
  then turn to methods that use bias potentials, including umbrella
  sampling and metadynamics.  We next consider parallel tempering and
  replica-exchange methods.  We conclude with a brief presentation of
  some combination methods.  }


\newcommand{\uu}[1]{{\boldsymbol #1}}
\newcommand{\cfa}[1]{{\color{black}#1}}
\newcommand{\gb}[1]{{\color{black}#1}}

\def\bb{\uu{b}}

\def\xb{\uu{x}}

\def\zb{\uu{z}}

\def\etab{\uu{\eta}}
\def\thetab{\uu{\theta}}
\def\lambdab{\uu{\lambda}}

\begin{document}

\section{Introduction}

The purpose of molecular dynamics (MD) is to compute the positions and velocities of a set
of interacting atoms at the present time instant given these
quantities one time increment in the past.  Uniform sampling from the
discrete trajectories one can generate using MD has long been seen as
synonymous with sampling from a statistical-mechanical ensemble; this
just expresses our collective wish that the ergodic hypothesis holds
at finite times.  Unfortunately, most MD trajectories are not ergodic
and leave many relevant regions of configuration space unexplored.
This stems from the separation of high-probability ``metastable''
regions by low-probability ``transition'' regions and the inherent
difficulty of sampling a 3$N$-dimensional space by embedding into it a
one-dimensional dynamical trajectory.

This review concerns a selection of methods to use MD simulation to
enhance the sampling of configuration space.  A central concern with
any enhanced sampling method is guaranteeing that the statistical
weights of the samples generated are known and correct (or at least
correctable) while simultaneously ensuring that as much of the
relevant regions of configuration space are sampled.  Because of the
tight relationship between probability and free energy, many of these
methods are known as ``free-energy'' methods.  To be sure, there are a
large number of excellent reviews of free-energy methods in the
literature
(e.g.,~\cite{Kollman1993,Trzesniak2007,Vanden-Eijnden2009,dellago2009transition,Christ2010}).
The present review is in no way intended to be as comprehensive as
these; as the title indicates, we will mostly focus on enhanced
sampling methods of three flavors: tempering, metadynamics, and
temperature-acceleration.  Along the way, we will point out important
related methods, but in the interest of brevity we will not spend much
time explaining these.  The methods we have chosen to focus on reflect
our own preferences to some extent, but they also represent popular
and growing classes of methods that find ever more use in biomolecular
simulations and beyond.

We divide our review into three main sections.  In the first, we
discuss enhanced sampling approaches that rely on {\it collective
  variable biasing}.  These include the historically important methods
of thermodynamic integration and umbrella sampling, and we pay
particular attention to the more recent approaches of the
adaptive-biasing force algorithm, temperature-acceleration, and
metadynamics.  In the second section, we discuss approaches based on
{\it tempering}, which is dominated by a discussion of the parallel
tempering/replica exchange approaches.  In the third section, we
briefly present some relatively new methods derived from either
\gb{collective-variable}-based or tempering-based approaches, or their combinations.

\section{Approaches Based on Collective-Variable Biasing}\label{sec:cv}
\subsection{Background: Collective Variables and Free Energy}
For our purposes, the term ``collective variable'' or CV refers to any
multidimensional function $\thetab$ of 3$N$-dimensional atomic
configuration $\xb\equiv\left(x_i|i=1\dots 3N\right)$.  The functions
$\theta_1(\xb)$, $\theta_2(\xb)$,$\dots$,$\theta_M(\xb)$ map
configuration $\xb$ onto an $M$-dimensional CV space
$\zb\equiv\left(z_j|j=1\dots M\right)$, where usually $M\ll 3N$.  
At equilibrium, the probability of observing the system at CV-point
$\zb$ is the weight of all configurations $\xb$
which map to $\zb$:
\begin{equation}\label{eq:p_of_zb}
P(\zb) = \left<\delta[\thetab(\xb)-\zb]\right>,
\end{equation}
The Dirac delta
function picks out only those configurations for which the CV
$\thetab(\xb)$ is $\zb$, and $\left<\cdot\right>$ denotes averaging
its argument over the equilibrium probability distribution of $\xb$.
The probability can
be expressed as a {\it free energy}:
\begin{equation}
\label{eq:restricted_free_energy} %
F(\zb) = -k_BT \ln\left<\delta[\thetab(\xb)-\zb]\right>.
\end{equation}
Here, $k_B$ is Boltzmann's constant and $T$ is temperature.

Local minima in $F$ are metastable equilibrium states.  $F$ also
measures the energetic cost of a maximally efficient (i.e.,
reversible) transition from one region of CV space to another.  If,
for example, we choose a CV space such that two well-separated regions
define two important allosteric states of a given protein, we could
perform a free-energy calculation to estimate the change in free
energy required to realize the conformational transition.  Indeed, the
promise of being able to observe with atomic detail the transition
states along some pathway connecting two distinct states of a
biomacromolecule is strong motivation for exploring these transitions
with CV's.

Given the limitations of standard MD, how does one ``discover'' such
states in a proposed CV space?  A perfectly ergodic (infinitely long)
MD trajectory would visit these minima much more frequently than it
would the intervening spaces, allowing one to tally how often each
point in CV space is visited; normalizing this histogram into a
probability $P(\zb)$ would be the most straightforward way to compute
$F$ via Eq.~\ref{eq:restricted_free_energy}.  In all too many actual
cases, MD trajectories remain close to only one minimum (the one
closest to the initial state of the simulation) and only very rarely,
if ever, visit others.  In the CV sense, we therefore speak of
standard MD simulations failing to overcome {\it barriers} in free
energy.  ``Enhanced sampling'' in this context refers then to methods
by which free-energy barriers in a chosen CV space are surmounted to
allow as broad as possible an extent of CV space to be explored and
statistically characterized with limited computational resources.

In this section, we focus on methods of enhanced sampling of CV's
based on MD simulations that are directly biased on those CV's; that
is, we focus on methods in which an investigator must identify the
CV's of interest as an input to the calculation.  We have chosen to
limit discussion to two broad classes of biasing: those whose
objective is direct computation of the gradient of the free energy
$(\partial F/\partial\zb)$ at local points throughout CV space, and
those in which non-Boltzmann sampling with bias potentials is used to
force exploration of otherwise hard-to-visit regions of CV space.  The
canonical methods in these two classes are {\it
  thermodynamic integration} and {\it umbrella sampling},
respectively, and a discussion of these two methods sets the stage for
discussion of three relatively modern variants: the Adaptive-Biasing
Force Algorithm~\cite{Darve2008}, Temperature-Accelerated
MD~\cite{Maragliano2006} and Metadynamics~\cite{Laio2002}.

\subsection{Gradient Methods:  Blue-Moon Sampling, Adaptive-Biasing Force Algorithm, and Temperature-Accelerated MD}

\subsubsection{Overview: Thermodynamic Integration}

Naively, one way to have an MD system visit a hard-to-reach point
$\zb$ in CV space is simply to create a realization of the
configuration $\xb$ at that point (i.e., such that $\thetab(\xb) =
\zb$).  This is an inverse problem, since the number of degrees of
freedom in $\xb$ is usually much larger than in $\zb$.  One way to
perform this inversion is by introducing external
forces that guide the configuration to the desired point from some
easy-to-create initial state; both targeted MD~\cite{Schlitter1993}
and steered MD~\cite{Grubmueller1996} are ways to do this.  Of course,
one would like MD to explore CV space in the vicinity of $\zb$, so
after creating the configuration $\xb$, one would just let it run.
Unfortunately, this would likely result in the system drifting away
from $\zb$ rather quickly, and there would be no way from such
calculations to estimate the likelihood of observing an unbiased long
MD simulation visit $\zb$.  But there is information in the fact that
the system drifts away; if one knows {\it on average} which direction
and how strongly the system would like to move if initialized at
$\zb$, this would be a measure of negative gradient of the free
energy, $-(\partial F/\partial\zb)$, or the ``mean force''.  We have
then a glimpse of a three-step method to compute $F$ (i.e., the
statistics of CV's) over a meaningfully broad extent of CV space:
\begin{enumerate}
\item 
visit a select number of local points in that space, and at each one,
\item compute the mean force, then
\item use numerical integration to reconstruct $F$ from these local
  mean forces; formally expressed as
\begin{equation}\label{eq:ti}
F(\zb)-F(\zb_0) = \int_{\zb_0}^{\zb} \left(\frac{\partial F}{\partial\zb}\right) d\zb
\end{equation}
\end{enumerate}
Inspired by Kirkwood's original suggestion involving switching
parameters~\cite{Kirkwood1935}, such an approach is generally referred
to as ``thermodynamic integration'' or TI.  TI allows us to
reconstruct the statistical weights of any point in CV space by
accumulating information on the gradients of free energy at selected
points.

\subsubsection{Blue-Moon Sampling}

The discussion so far leaves open the correct way to compute the local
free-energy gradients.  A gradient is a local quantity, so a natural
choice is to compute it from an MD simulation localized at a point in
CV space by a constraint.  Consider a long MD simulation with a
holonomic constraint fixing the system at the point $\zb$.  Uniform
samples from this constrained trajectory $\xb(t)$ then represent {\it
  an} ensemble at fixed $\zb$ over which the averaging needed to
convert gradients in potential energy to gradients in free energy
could be done.  However, this constrained ensemble has the undesired
property that the velocities $\dot\thetab(\xb)$ are zero.  This is a
bit problematic because virtually none of the samples plucked from a
long unconstrained MD simulation (as is implied by
Eq.~\ref{eq:p_of_zb}), would have $\dot\thetab=0$, and $\dot\thetab=0$
acts as a set of $M$ unphysical constraints on the system velocities
$\dot\xb$, since $\dot\theta_j = \sum_i (\partial\theta_j/\partial
x_i)\dot x_i$.  Probably the best-known example of a method to correct
for this bias is the so-called ``blue-moon'' sampling
method~\cite{Carter1989,Sprik1998,Ciccotti2004,Ciccotti2005} or the
constrained ensemble method~\cite{denOtter1998,Schlitter2003}.  The
essence of the method is a decomposition of free energy gradients into
components along the CV gradients and thermal components orthogonal to
them:
\begin{equation}\label{eq:blue-moon}
\frac{\partial F}{\partial z_j} = \left<\bb_j(\xb)\cdot\nabla V(\xb) 
- k_BT\nabla\cdot \bb_j(\xb)\right>_{\thetab(\xb)=\zb}
\end{equation}
where $\left<\cdot\right>_{\thetab(\xb)=\zb}$ denotes averaging across
samples drawn uniformly from the MD simulation constrained at
${\thetab(\xb)=\zb}$, and the $\bb_j(\xb)$ is the vector field
orthogonal to the gradients of every component $k$ of $\thetab$ for
$k\ne j$:
\begin{equation}
\bb_j(\xb)\cdot\nabla\theta_k(\xb) = \delta_{jk}
\end{equation}
where $\delta_{jk}$ is the Kroenecker delta.  (For brevity, we have
omitted the consideration of holonomic constraints other than that on
the CV; the reader is referred to the paper by Ciccotti et al. for
details~\cite{Ciccotti2005}.)  The vector fields $\bb_j$ for each
$\theta_j$ can be constructed by orthogonalization.  The first term in
the angle brackets in Eq.~\ref{eq:blue-moon} implements the chain rule
one needs to account for how energy $V$ changes with $\zb$ through all
the ways $\zb$ can change with $\xb$.  The second term corrects for
the thermal bias imposed by the constraint.

Although nowhere near exhaustive, below is a listing of common types
of problems to which blue-moon sampling has been applied with some
representative examples:
\begin{enumerate}
\item sampling conformations of small flexible molecules and
  peptides~\cite{Depaepe1993,Zhao2008,Kim2009}
\item environmental effects on covalent bond formation/breaking
  (usually in combination with {\it ab initio}
  MD)~\cite{Hytha2001,Fois2004,Ivanov2005,Stubbs2005,Trinh2009,LiuP2010,Bucko2010}
\item solvation and non-covalent binding of small molecules in
  solvent~\cite{Paci1994,Sa2006,Mugnai2007,Chunsrivirot2011,Sato2012}
\item protein dimerization~\cite{Sergi2002,Maragliano2004}

\end{enumerate}

\subsubsection{The Adaptive Biasing Force Algorithm}

The blue-moon approach requires multiple independent constrained MD
simulations to cover the region of CV space in which one wants
internal statistics.  The care taken in choosing these quadrature
points can often dictate the accuracy of the resulting free energy
reconstruction.  It is therefore sometimes advantageous to consider
ways to avoid having to choose such points ahead of time, and adaptive
methods attempt to address this problem.  One example is the
adaptive-biasing force (ABF) algorithm of Darve et
al.~\cite{Darve2001,Darve2008} The essence of ABF is two-fold: (1)
recognition that external bias forces of the form
$\nabla_\xb\theta_j\left(\partial F/\partial z_j\right)$ for $j=1\dots
M$ exactly oppose mean forces and should lead to more uniform sampling
of CV space, and (2) that these bias forces can be converged upon
adaptively during a single unconstrained MD simulation.

The first of those two ideas is motivated by the fact that ``forces''
that keep normal MD simulations effectively confined to free energy
minima are mean forces on the collective variables projected onto the
atomic coordinates, and balancing those forces against their exact
opposite should allow for thermal motion to take the system out of
those minima.  The second idea is a bit more subtle; after all, in a
running MD simulation with no CV constraints, the constrained ensemble
expression for the mean force (Eq.~\ref{eq:blue-moon}) does not
directly apply, because a constrained ensemble is not what is being
sampled.  However, Darve et al. showed how to relate these ensembles
so that the samples generated in the MD simulation could be used to
build mean forces~\cite{Darve2001}.  Further, they showed using a
clever choice of the fields of Eq.~\ref{eq:blue-moon} an equivalence
between ($i$) the spatial gradients needed to computed forces, and
($ii$) time-derivatives of the CV's~\cite{Darve2008}:
\begin{equation}\label{eq:abf}
\frac{\partial F}{\partial z_i} = -k_BT\left<\frac{d}{dt}\left(M_\theta\frac{d\theta_i}{dt}\right)\right>_{\thetab=\zb}
\end{equation}
where $M_\theta$ is the transformed mass matrix given by
\begin{equation}\label{eq:abf_mm}
M_\theta^{-1} = J_\theta M^{-1} J_\theta
\end{equation}
where $J_\theta$ is the $M\times 3N$ matrix with elements
$\partial\theta_i/\partial x_j$ ($i = 1\dots M$, $j = 1\dots 3N$), and
$M$ is the diagonal matrix of atomic masses.  Eq.~\ref{eq:abf_mm} is
the result of a particular choice for the fields $\bb_j(\xb)$.  This
reformulation of the instantaneous mean forces computed on-the-fly
makes ABF exceptionally easy to implement in most modern MD packages.
Darve et al. present a clear demonstration of the ABF algorithm in
\cfa{a\ } pseudocode~\cite{Darve2008} that attests to this fact.

ABF has found rather wide application in CV-based free energy calculations
in recent years.  Below is \cfa{a\ } representative sample of some types
of problems subjected to ABF calculations in the recent literature:
\begin{enumerate}
\item
Peptide backbone angle sampling~\cite{Fogolari2011,Faller2013};
\item Nucleoside~\cite{Wei2011}, protein~\cite{Vivcharuck2011} and
  fullerene~\cite{Kraszewski2011,Kraszewski2012} insertion into a
  lipid bilayer;
\item Interactions of small molecules with polymers in
  water~\cite{Liu2010,Caballero2013};
\item Molecule/ion transport through protein
  complexes~\cite{Wilson2011,Cheng2012,%
    Wang2012,Tillman2013} and DNA superstructures~\cite{Akhshi2012};
\item Calculation of octanol-water partition
  coefficients~\cite{Kamath2012,Bhatnagar2012};
\item Large-scale protein conformational
  changes~\cite{Wereszczynski2012};
\item Protein-nanotube~\cite{Jana2012} and
  nanotube-nanotube~\cite{Uddin2010} association.
\end{enumerate}

\subsubsection{Temperature-Accelerated Molecular Dynamics}

Both blue-moon sampling and ABF are based on statistics in the
constrained ensemble.  However, estimation of mean forces need not
only use this ensemble.  One can instead relax the constraint and work
with a ``mollified'' version of the free energy:
\begin{equation}
\label{eq:mollified_free_energy}
F_\kappa(\zb) = -k_BT\ln\left<\delta_\kappa\left[\thetab(\xb)-\zb\right]\right>
\end{equation}
where $\delta_\kappa$ refers to the Gaussian (or ``mollified delta function''):
\begin{equation}
\label{eq:mollified_delta_function}
\delta_\kappa = \sqrt{\frac{\beta\kappa}{2\pi}} \exp\left[-\frac12\beta\kappa\left|\thetab(\xb)-\zb\right|^2\right],
\end{equation}
where $\beta$ is just shorthand for $1/k_BT$.  Since
$\lim_{\beta\kappa\rightarrow\infty}\delta_\kappa = \delta$, we know
that $\lim_{\beta\kappa\rightarrow\infty}F_\kappa = F$.  One way to
view this Gaussian is that it ``smoothes out'' the true free energy to
a tunable degree; the factor $1/\sqrt{\beta\kappa}$ is a length-scale
in CV space below which details are smeared.

Because the Gaussian has continuous gradients, it can be used directly
in an MD simulation. Suppose we have a CV space $\thetab(\xb)$, and we
extend our MD system to include variables $\zb$ such that the combined
set $(\xb,\zb)$ obeys the following extended potential:
\begin{equation}\label{eq:tamd_potential}
U(\xb,\zb) = V(\xb) + \sum_{j=1}^M \frac12\kappa\left|\theta_j(\xb)-z_j\right|^2
\end{equation}
where $V(\xb)$ is the interatomic potential, and $\kappa$ is a
constant.  Clearly, if we fix $\zb$, then the resulting free energy is
to within an additive constant the mollified free energy of
Eq.~\ref{eq:mollified_free_energy}.  (The additive constant is related
to the prefactor of the mollified delta function and has nothing to do
with the number of CV's.) Further, we can directly express the
gradient of this mollified free energy with respect to $\zb$:~\cite{Kaestner2009}
\begin{equation}\label{eq:tamd_free_energy_gradient}
\nabla_\zb F_\kappa = -\left<\kappa\left[\thetab(\xb)-\zb\right]\right>
\end{equation}
This suggests that, instead of using constrained ensemble MD to
accumulate mean forces, we could work in the {\em restrained} ensemble
and get very good approximations to the mean force.  By
``restrained'', we refer to the fact that the term giving rise to the
mollified delta function in the configurational integral is
essentially a harmonic restraining potential with a ``spring
constant'' $\kappa$.  In this restrained-ensemble approach, no
velocities are held fixed, and the larger we choose $\kappa$ the more
closely we can approximate the true free energy.  Notice however that
large values of $\kappa$ could lead to numerical instabilities in
integrating equations of motion, and a balance should be found.  (In
practice, we have found that for CV's with dimensions of length,
values of $\kappa$ less than about 1,000 kcal/mol/\AA$^2$ can be
stably handled, and values of around 100 kcal/mol/\AA$^2$ are
typically adequate.)

Temperature-accelerated MD (TAMD)~\cite{Maragliano2006} takes
advantage of the restrained-ensemble approach to directly evolve the
variables $\zb$ in such a way to accelerate the sampling of CV space.
First, consider how the atomic variables $\xb$ evolve under the
extended potential (assuming Langevin dynamics):
\begin{equation}\label{eq:tamd_atomic}
m_i\ddot{x}_i = -\frac{\partial V(\xb)}{\partial x_i} -
\kappa\sum_{j=1}^m\left[\theta_j(\xb)-z_j\right]
\frac{\partial\theta_j(\xb)}{\partial x_i} - 
\gamma m_i\dot{x_i} + \eta_i(t;\beta)
\end{equation}
Here, $m_i$ is the mass of $x_i$, $\gamma$ is the friction coefficient
for the Langevin thermostat, and $\etab$ is the thermostat white noise
satisfying the fluctuation-dissipation theorem at physical temperature
$\beta^{-1}$:
\begin{equation}
 \label{eq:noisedef}
 \left<\eta_i(t;\beta)\eta_j(t';\beta)\right> = \beta^{-1}
  \gamma m_i \delta_{ij}\delta(t-t')
\end{equation}
Key to TAMD is that the $\zb$ are treated as slow variables that
evolve according to their own equations of motion, which here we take
as diffusive (though other choices are
possible~\cite{Maragliano2006}):
\begin{equation}
\label{eq:slow_dynamics}
\bar{\gamma}\bar{m}_j\dot{z}_j = \kappa\left[\theta_j(\xb)-z_j\right] 
+ \xi_j(t;\bar \beta).
\end{equation}
Here, $\bar\gamma$ is a fictitious friction, $\bar{m}_j$ is a mass,
and the first term on the right-hand side represents the instantaneous
force on variable $z_j$, and the second term represents thermal noise
at the fictitious thermal energy $\bar\beta^{-1} \not = \beta^{-1}$.

The advantage of TAMD is that if (1) $\bar\gamma$ is chosen
sufficiently large so as to guarantee that the slow variables indeed
evolve slowly relative to the fundamental variables, {\em and} (2)
$\kappa$ is sufficiently large such that
$\thetab(\xb(t))\approx\zb(t)$ at any given time, then the force
acting on $\zb$ is approximately equal to minus the gradient of the
free energy
(Eq.~\ref{eq:tamd_free_energy_gradient})~\cite{Maragliano2006}. This
is because the MD integration repeatedly samples
$\kappa\left[\thetab(\xb)-\zb\right]$ for an essentially fixed (but
actually very slowly moving) $\zb$, so $\zb$ evolution effectively
feels these samples as a mean force.  In other words, the dynamics of
$\zb(t)$ is effectively
\begin{equation}
\bar{\gamma}\bar m_j \dot{z}_j = -\frac{\partial F(\zb)}{\partial z_j} 
+ \xi_j(t;\bar \beta).
\end{equation}
This shows that the $\zb$-dynamics describes an equilibrium
constant-temperature ensemble at {\em fictitious} temperature
$\bar\beta^{-1}$ acted on by the ``potential'' $F(\zb)$, which is the
free energy evaluated at the {\em physical} temperature $\beta^{-1}$.
That is, under TAMD, $\zb$ conforms to a probability distribution of
the form $\exp\left[-\bar\beta F(\zb;\beta)\right]$, whereas under
normal MD it would conform to $\exp\left[-\beta F(\zb;\beta)\right]$.
The all-atom MD simulation (at $\beta$) simply serves to approximate
the {\em local gradients} of $F(\zb)$.  Sampling is enhanced by taking
$\bar\beta^{-1}>\beta^{-1}$, which has the effect of attenuating the
ruggedness of $F$.  TAMD therefore can accelerate a trajectory
$\zb(t)$ through CV space by increasing the likelihood of visiting
points with relatively low physical Boltzmann factors.  This borrows
directly from the main idea of adiabatic free-energy
dynamics~\cite{Rosso2002} (AFED), in that one deliberately makes some
variables hot (to overcome barriers) but slow (to keep them
adiabatically separated from all other variables).  In TAMD, however,
the use of the mollified free energy means no cumbersome variable
transformations are required. (The authors of AFED refer to TAMD as
``driven''-AFED, or d-AFED~\cite{AbramsJ2008}.)  It is also worth
mentioning in this review that TAMD borrows heavily from an early
version of metadynamics~\cite{Iannuzzi2003}, which was formulated as a
way to evolve the auxiliary variables $\zb$ on a mollified free
energy.  However, unlike metadynamics (which we discuss below in
Sec.~\ref{sec:metadynamics}), there is no history-dependent bias in
TAMD.

Unlike TI, ABF, and the methods of umbrella sampling and metadynamics
discussed in the next section, TAMD is not a method for direct
calculation of the free energy.  Rather, it is a way to overcome free
energy barriers in a chosen CV space quickly without visiting
irrelevant regions of CV space.  (However, we discuss briefly a method
in Sec.~\ref{sec:otf_fep} in which TAMD gradients are used in a spirit
similar to ABF to reconstruct a free energy.)  That is, we consider
TAMD a way to efficiently explore relevant regions CV space that are
practically inaccessible to standard MD simulation.  It is also worth
pointing out that, unlike ABF, TAMD does not operate by opposing the
natural gradients in free energy, but rather by using them to guide
accelerated sampling.  ABF can only use forces in locations in CV
space the trajectory has visited, which means nothing opposes the
trajectory going to regions of very high free energy.  However, under
TAMD, an acceleration of $\bar\beta^{-1}$= 6 kcal/mol on the CV's will
greatly accelerate transitions over barriers of 6-12 kcal/mol, but
will still not (in theory) accelerate excursions to regions requiring
climbs of hundreds of kcal/mol.  TAMD and ABF have in common the
ability to handle rather high-dimensional CV's.

Although it was presented theoretically in 2006~\cite{Maragliano2006},
TAMD was not applied directly to large-scale MD until much
later~\cite{Abrams2010}.  Since then, there has been growing interest
in using TAMD in a variety of applications requiring enhanced
sampling:
\begin{enumerate}
\item TAMD-enhanced flexible fitting of all-atom protein and RNA
  models into low-resolution electron microscopy density
  maps~\cite{Vashisth2012c,Vashisth2013a};
\item Large-scale (interdomain) protein conformational
  sampling~\cite{Abrams2010,Vashisth2012a,Hu2012};
\item Loop conformational sampling in proteins~\cite{Vashisth2012b};
\item Mapping of diffusion pathways for small molecules in globular
  proteins~\cite{Maragliano2010,Lapelosa2013};
\item Vacancy diffusion~\cite{Geslin2013};
\item Conformational sampling and packing in dense polymer
  systems~\cite{Lucid2013}.
\end{enumerate}

Finally, we mention briefly that TAMD can be used as a quick way to
generate trajectories from which samples can be drawn for subsequent
mean-force estimation for later reconstruction of a multidimensional
free energy; this is the essence of the single-sweep
method~\cite{Maragliano2008}, which is an efficient means of computing
multidimensional free energies.  Rather than using straight numerical
TI, single sweep posits the free energy as a basis function expansion
and uses standard optimization methods to find the expansion
coefficients that best reproduce the measured mean forces.
Single-sweep has been used to map diffusion pathways of CO and H$_2$O
in myoglobin~\cite{Maragliano2010,Lapelosa2013}.

\subsection{Bias Potential Methods:  Umbrella Sampling and Metadynamics}

\subsubsection{Overview: Non-Boltzmann Sampling}

In the previous section, we considered methods that achieve enhanced
sampling by using mean forces: in TI, these are integrated to
reconstruct a free energy; in ABF, these are built on-the-fly to drive
uniform CV sampling; and in TAMD, these are used on-the-fly to guide
accelerated evolution of CV's.  In this section, we consider methods
that achieve enhanced sampling by means of controlled bias potentials.
As a class, we refer to these as {\em non-Boltzmann sampling} methods.

Non-Boltzmann sampling is generally a way to derive statistics on a
system whose energetics differ from the energetics used to perform the
sampling.  Imagine we have an MD system with bare interatomic
potential $V(\xb)$, and we add a bias $\Delta V(\xb)$ to arrive at a
biased total potential:
\begin{equation}\label{eq:nbs_biased_total_potential}
V_b(\xb) = V(\xb) + \Delta V(\xb)
\end{equation}
The statistics on the CV's on this biased potential are then given as
\begin{align}\label{eq:nbs_biased_statistics}
P_b(\zb) & = \frac{\displaystyle \int\! d\xb\ e^{-\beta V_0(\xb)} e^{-\beta \Delta V(\xb)} \delta\left[\thetab(\xb)-\zb\right]}{\displaystyle\int\! d\xb\ e^{-\beta V_0(\xb)} e^{-\beta \Delta V(\xb)}}\nonumber\\
& = \frac{\displaystyle\displaystyle \int\! d\xb\ e^{-\beta V_0(\xb)} e^{-\beta \Delta V} \delta\left[\thetab(\xb)-\zb\right]}{\displaystyle\int\! d\xb\ e^{-\beta V_0(\xb)}}\frac{\int\! d\xb\ e^{-\beta V_0(\xb)}}{\displaystyle\int\! d\xb\ e^{-\beta V_0(\xb)} e^{-\beta \Delta V(\xb)}}\nonumber\\
& = \frac{\left<e^{-\beta\Delta V(\xb)}\delta\left[\thetab(\xb)-\zb\right]\right>}{\left<e^{-\beta\Delta V(\xb)}\right>}
\end{align}
where $\left<\cdot\right>$ denotes ensemble averaging on the unbiased
potential $V(\xb)$. Further, if we take the bias potential $\Delta V$
to be explicitly a function only of the CV's $\thetab$, then it
becomes invariant in the averaging of the numerator thanks to the
delta function, and we have
\begin{equation}\label{eq:nbs_biased_pullout}
P_b(\xb) = \frac{e^{-\beta \Delta V(\zb)} \left<\delta\left[\thetab(\xb)-\zb\right]\right>}{\left<e^{-\beta\Delta V\left[\theta(\xb)\right]}\right>}
\end{equation}
Finally, since the unbiased statistics are $P(\zb) = \left<\delta\left[\thetab(\xb)-\zb\right]\right>$, we arrive at
\begin{equation}\label{eq:nbs_biased_to_unbiased}
P(\zb) = P_b(\zb) e^{\beta \Delta V(\zb)} \left<e^{-\beta\Delta V\left[\theta(\xb)\right]}\right>
\end{equation}
\cfa{Taking\ } samples from an ergodic MD simulation \cfa{on\ } the
biased potential $V_b$, Eq.~\ref{eq:nbs_biased_to_unbiased} provides
the recipe for reconstructing the statistics the CV's {\em would}
present were they generated using the {\em unbiased} potential $V$.
However, the probability $P(\zb)$ \cfa{is\ } implicit in this
equation, because
\begin{equation}\label{eq:nbs_implicit_thing}
\left<e^{-\beta\Delta V}\right> = \int d\zb P(\zb) e^{-\beta\Delta V\left[\thetab(\xb)\right]}
\end{equation}
This is not really a problem, since we can treat
$\left<e^{-\beta\Delta V}\right>$ as a constant we can get from
normalizing $P_b(\zb) e^{\beta \Delta V(\zb)}$.

How does one choose $\Delta V$ so as to enhance the sampling of CV
space?  Evidently, from the standpoint of non-Boltzmann sampling, the
closer the bias potential is to the negative free energy $-F(\zb)$,
the more uniform the sampling of CV space will be.  To wit: if $\Delta
V\left[\thetab(\xb)\right] = -F\left[\thetab(\xb)\right]$, then
$e^{\beta \Delta V(\zb)}=e^{-\beta F(\zb)}=P(\zb)$, and
Eq.~\ref{eq:nbs_biased_to_unbiased} can be inverted for $P_b$ to yield
\begin{equation}\label{eq:nbs_opt}
P_b(\zb)=\frac{1}{\left<e^{\beta F(\zb)}\right>}=\frac1{\displaystyle\int d\zb P(\zb) e^{\beta F(\zb)}}=\frac1{\displaystyle\int d\zb e^{-\beta F}e^{\beta F}} = \frac1{\displaystyle\int d\zb}
\end{equation}
So we see that taking the bias potential to be the negative free
energy makes all states $\zb$ in CV space equiprobable.  This is
indeed the limit to which ABF strives by applying negative mean
forces, for example~\cite{Darve2008}.

We usually do not know the free energy ahead of time; if we did, we
would already know the statistics of CV space and no enhanced sampling
would be necessary.  Moreover, perfectly uniform sampling of the
entire CV space is usually far from necessary, since most CV spaces
have many irrelevant regions that should be ignored.  And in reference
to the mean-force methods of the last section, uniform sampling is
likely not necessary to achieve accurate mean force values; how good
an estimate of $\nabla F$ is at some point $\zb_0$ should not depend
on how well we sampled at some other point $\zb_1$.  Yet achieving
uniform sampling is an idealization since, if we do, this means we
know the free energy.  We now consider two other biasing methods that
aim for this ideal, either in relatively small regions of CV space
using fixed biases, or over broader extents using adaptive biases.
  
\subsubsection{Umbrella Sampling}

Umbrella sampling is the standard way of using non-Boltzmann sampling
to overcome free energy barriers.  In its
debut~\cite{torrie1977nonphysical}, umbrella sampling used a function
$w(\xb)$ that weights hard-to-sample configurations, equivalent to
adding a bias potential of the form
\begin{equation}\label{eq:umbrella_sampling_bias}
\Delta V(\xb) = -k_BT\ln w(\xb).
\end{equation}
$w$ is found by trial-and-error such that configurations that are easy
to sample on the unbiased potential are still easy to sample; that is,
$w$ acts like an ``umbrella'' covering both the easy- and
hard-to-sample regions of configuration space.  Nearly always, $w$ is
an explicit function of the CV's, $w(\xb) = W[\thetab(\xb)]$.

Coming up with the umbrella potential that would enable exploration of
CV space with a single umbrella sampling simulation that takes the
system far from its initial point is not straightforward.  Akin to TI,
it is therefore advantageous to combine results from
several independent trajectories, each with its own umbrella potential
that localizes it to a small volume of CV space that overlaps with
nearby volumes.  The most popular way to combine the statistics of
such a set of independent umbrella sampling runs is the
weighted-histogram analysis method (WHAM)~\cite{kumar1992weighted}.

To compute statistics of CV space using WHAM, one first chooses the
points in CV space that define the little local neighborhoods, or
``windows'' to be sampled and chooses the bias potential used to
localize the sampling.  Not knowing how the free energy changes in CV
space makes the first task somewhat challenging, since more densely
packed windows are preferred in regions where the free energy changes
rapidly; however, since the calculations are independent, more can be
added later if needed.  A convenient choice for the bias potential is
a simple harmonic spring that tethers the trajectory to a reference
point $\zb_i$ in CV space:
\begin{equation}\label{eq:wham_harmonic}
\Delta V_i(\xb) = \frac12\kappa\left|\thetab(\xb)-\zb_i\right|^2
\end{equation}
which means the dynamics of the atomic variables $\xb$ are identical
to Eq.~\ref{eq:tamd_atomic} at fixed $\zb =\zb_i$.  The points
$\left\{\zb_i\right\}$ and the value of $\kappa$ (which may be
point-dependent) must be chosen such that $\thetab\left[\xb(t)\right]$
from any one window's trajectory makes excursions into the window of
each of its nearest neighbors in CV space.

Each window-restrained trajectory is directly histogrammed to yield
apparent (i.e., biased) statistics on $\thetab$; let us call the
biased probability in the $i$th window $P_{b,i}(\zb)$.
Eq.~\ref{eq:nbs_biased_to_unbiased} again gives the recipe to
reconstruct the unbiased statistics $P_i(\zb)$ for $\zb$ in the window
of $\zb_i$:
\begin{equation}\label{eq:biased_us}
P_i(\zb) = P_{b,i}(\zb) e^{\frac12\beta\kappa\left|\zb-\zb_i\right|^2}\left<e^{-\beta\frac12\kappa\left|\thetab(\xb)-\zb_i\right|^2}\right>
\end{equation}
We could use Eq.~\ref{eq:biased_us} directly assuming the biased MD
trajectory is ergodic, but we know that regions far from the reference
point will be explored very rarely and thus their free energy would be
estimated with large uncertainty.  This means that, although we can
use sampling to compute $P_{b,i}$ {\em knowing} it effectively
vanishes outside the neighborhood of $\zb_i$, we cannot use sampling
to compute
$\left<e^{-\beta\frac12\kappa\left|\thetab(\xb)-\zb_i\right|^2}\right>$.

WHAM solves this problem by renormalizing the probabilities in each
window into a single composite probability.  Where there is overlap
among windows, WHAM renormalizes such that the statistical variance of
the probability is minimal.  That is, it treats the factor
$\left<e^{-\beta\frac12\kappa\left|\thetab(\xb)-\zb_i\right|^2}\right>$
as an undetermined constant $C_i$ for each window, and solves for
specific values such that the composite unbiased probability $P(\zb)$
is continuous across all overlap regions with minimal statistical
error.  An alternative to WHAM, termed ``umbrella integration'',
solves the problem of renormalization across windows by constructing
the composite mean force~\cite{Kaestner2005,Kaestner2011}.

The literature on umbrella sampling is vast (by simulation standards),
so we present here a very condensed listing of some of its more recent
application areas with representative citations:
\begin{enumerate}
\item Small molecule conformational sampling~\cite{Schaefer1998,Banavali2002,Cruz2011,Islam2011};
\item Protein-folding~\cite{Young1996,Sheinerman1998,Bursulaya1999} and large-scale protein conformational sampling~\cite{Rick1998,Allen2003,Shams2012,Yildirim2013};
\item Protein-protein/peptide-peptide interactions~\cite{Masunov2003,Tarus2005,Makowski2011,Casalini2011,Wanasundara2011,Zhang2011,Periole2012,Vijayaraj2012,Mahdavi2013};
\item DNA conformational changes~\cite{Banavali2005} and DNA-DNA interactions~\cite{Giudice2003,Matek2012,Bagai2013};
\item Binding and association free-energies~\cite{Czaplewski2000,Lee2006,Peri2011,St-Pierre2011,Rashid2012,Chen2012,Wilhelm2012,Louet2012,Zhang2012,Kessler2012,Mascarenhas2013};
\item Adsorption on and permeation through lipid bilayers~\cite{MacCallum2006,Tieleman2006,Kyrychenko2011,Lemkul2011,Paloncyova2012,Samanta2012,Grafmueller2013,Cerezo2013,Tian2013,Karlsson2013};
\item Adsorption onto inorganic surfaces/interfaces~\cite{Euston2011,Doudou2012};
\item Water ionization~\cite{Pomes1998,Jagoda-Cwiklik2011};
\item Phase transitions~\cite{Calvo2011,Sharma2012};
\item Enzymatic mechanisms~\cite{Ridder2002,Kaestner2006,Wang2007,Ke2011,Yan2011,Mujika2012,Lonsdale2012,Rooklin2012,Lior-Hoffmann2012};
\item Molecule/ion transport through protein complexes~\cite{Crouzy2001,Allen2003b,Hub2008,Xin2011,Furini2011,Kim2011,Domene2012,Zhu2012} and other macromolecules~\cite{He2011,Nalaparaju2012}.
\end{enumerate}

\subsubsection{Metadynamics}\label{sec:metadynamics}

As already mentioned, one of the difficulties of the umbrella sampling
method is the choice and construction of the bias potential.  As we
already saw with the relationship among TI, ABF, and TAMD, an adaptive
method for building a bias potential in a running MD simulation may be
advantageous.  Metadynamics~\cite{Laio2002,Barducci2011} represents
just such a method.

\cfa{Metadynamics is rooted in the original idea of ``local
  elevation''~\cite{Huber1994}, in which a supplemental bias potential
  is progressively grown \gb{in the dihedral space of a molecule} to
  prevent \gb{it} from remaining in one region of configuration
  space.}  \gb{ However, at variance with metadynamics, local
  elevation does not provide any means to reconstruct the unbiased
  free-energy landscape and as such it is mostly aimed at fast
  generation of plausible conformers.  }

In metadynamics,
configurational variables $\xb$ evolve in response to a biased total
potential:
\begin{equation}\label{eq:metadynamics_total_potential}
V(\xb) = V_0(\xb) + \Delta V(\xb,t)
\end{equation}
where $V_0$ is the bare interatomic potential and $\Delta V(\xb,t)$ is
a time-dependent bias potential.  The key element of metadynamics is
that the bias is built as a sum of Gaussian functions centered on the
points in CV space already visited:
\begin{equation}\label{eq:metadynamics_bias}
\Delta V\left[\thetab(\xb),t\right] = w\sum_{\begin{array}{c}t^\prime=\tau_G,2\tau_G,\dots\\t^\prime<t\end{array}}\exp\left(-\frac{\left|\thetab\left[\xb(t)\right]-\thetab\left[\xb(t^\prime)\right]\right|^2}{2\delta\thetab^2}\right)
\end{equation}
Here, $w$ is the height of each Gaussian, $\tau_G$ is the size of the
time interval between successive Gaussian depositions, and
$\delta\thetab$ is the Gaussian width.  It has been first
empirically~\cite{laio2005assessing} then
analytically~\cite{bussi2006equilibrium} demonstrated that in the
limit in which the CV evolve according to a Langevin dynamics, the
bias indeed converges to the negative of the free energy, thus
providing an optimal bias to enhance transition events. Multiple
simulations can also be used to allow for a quicker filling of the
free-energy landscape~\cite{raiteri2006efficient}.

The difference between the metadynamics estimate of the free energy
and the true free energy can be shown to be related to the diffusion
coefficient of the collective variables and to the rate at which the
bias is grown.  A possible way to decrease this error as a simulation
progresses is to decrease the growth rate of the bias. Well-tempered
metadynamics~\cite{barducci2008well} used an optimized schedule to
decrease the deposition rate of bias by modulating the Gaussian
height:
\begin{equation}\label{eq:well_tempered_metadynamics_w}
w=\omega_0\tau_Ge^{-\frac{\Delta V(\thetab,t)}{k_B\Delta T}}
\end{equation}
Here, $\omega_0$ is the initial ``deposition rate'', measured Gaussian
height per unit time, and $\Delta T$ is a parameter that controls \cfa{the\ }
degree to which the biased trajectory makes excursions away from
free-energy minima.  It is possible to show that using well-tempered
metadynamics the bias does not converge to the negative of the
free-energy but to a fraction of it, thus resulting in sampling the
CVs at an effectively higher temperature $T+\Delta T$, where normal
metadynamics is recovered for $\Delta T\rightarrow \infty$.  We notice
that other deposition schedules can be used aimed, e.g., at maximizing
the number of round-trips in the CV space \cite{singh2011flux}.
Importantly, it is possible to recover equilibrium Boltzmann
statistics of {\em unbiased} collective variables from samples drawn
throughout a well-tempered metadynamics trajectory~\cite{bononmi2009};
it does not seem clear that one can do this from an ABF trajectory.
Finally, it is possible to tune the shape of the Gaussians on the fly
using schemes based on the geometric compression of the phase space or
on the variance of the CVs \cite{branduardi2012metadynamics}.

In the well-tempered ensemble, the parameter $\Delta T$ can be used to
tune the size of the explored region, in a fashion similar to the
fictitious temperature in TAMD.  So both TAMD and well-tempered
metadynamics can be used to explore {\em relevant} regions of CV space
while surmounting {\em relevant} free energy barriers.  However, there
are important distictions between the two methods.  First, the main
source of error in TAMD rests with how well mean-forces are
approximated, and adiabatic separation, realizable only when the
auxiliary variables $\zb$ never move, is the only way to guarantee
they are perfectly accurate.  In practical application, TAMD never
achieves perfect adiabatic separation.  In contrast, because the
deposition rate of decreases as a well-tempered trajectory
progresses, errors related to poor adiabatic separation are
progressively damped.  Second, as already mentioned, TAMD alone cannot
report the free energy, but it also is therefore not practically
limited by the dimensionality of CV space; multicomponent gradients
are just as accurately calculated in TAMD as are single-component
gradients.  Metadynamics, as a histogram-filling method, must
exhaustively sample a finite region around any point to know the free
energy and its gradients are correct, which can sometimes limit its
utility.

Metadynamics is a powerful method whose popularity continues to grow.
In either its original formulation or in more recent variants,
metadynamics has been employed successfully in several fields, some of
which we point out below with some representative examples:
\begin{enumerate}
\item Chemical reactions \cite{Iannuzzi2003,mcgrath2013atp};
\item Peptide backbone angle sampling \cite{mantz2009ensemble,leone2009mechanism,melis2009trans};
\item Protein folding \cite{bussi2006free,gangupomu2010,berteotti2011effect,granata2013characterization};
\item Protein aggregation \cite{baftizadeh2012multidimensional};
\item Molecular docking \cite{gervasio2005flexible,Soederhjelm2012locating,limongelli2013funnel};
\item Conformational rearrangement of proteins \cite{sutto2013effects};
\item Crystal structure prediction \cite{martonak2006crystal};
\item Nucleation and crystal growth \cite{trudu2006freezing,stack2011accurate};
\item and proton diffusion \cite{zhang2012water}.
\end{enumerate}

\subsection{Some Comments on Collective Variables}\label{sec:cv_comments}

\subsubsection{The Physical Fidelity of CV-Spaces}

Given a potential $V(\xb)$, any multidimensional CV $\thetab(\xb)$ has
a mathematically determined free energy $F(\zb)$, and in principle the
free-energy methods we describe here (and others) can use and/or
compute it.  However, this does not guarantee that $F$ is meaningful,
and a poor choice for $\thetab(\xb)$ can render the results of even
the most sophisticated free-energy methods useless for understanding
the nature of actual metastable states and the transitions among them.
This puts two major requirements on any CV space:
\begin{enumerate}
\item Metastable states and transition states must be unambiguously
  identified as {\em energetically} separate regions in CV space.
\item The CV space must not contain hidden barriers.
\end{enumerate}

The first of these may seem obvious: CV's are chosen to provide a
low-dimensional description of some important process, say a
conformational change or a chemical reaction or a binding event, and
one can't describe a process without being able to discriminate
states.  However, it is not always easy to find CV's that do this.
Even given representative configurations of two distinct metastable
states, standard MD from these two different initial configurations
may sample partially overlapping regions of CV space, making ambiguous
the assignation of an arbitrary configuration to a state.  It may be
in this case that the two representative configurations actually
belong to the same state, or that if there are two states, that no
matter what CV space is overlaid, the barrier separating them is so
small that, on MD timescales, they can be considered rapidly
exchanging substates of some larger state.

But a third possibility exists: the two MD simulations mentioned above
may in fact represent very different states. The overlap might just be
an artifact of neglecting to include one or more CV's that are truly
necessary to distinguish those states.  If there is a significant free
energy barrier along this neglected variable, an MD simulation will
not cross it, yet may still sample regions in CV space also sampled by
an MD simulation launched from the other side of this hidden barrier.
And it is even worse: if TI or umbrella sampling is used along a
pathway in CV space that neglects an important variable, the
free-energy barriers along that pathway might be totally meaningless.

Hidden barriers can be a significant problem in CV-based free-energy
calculations. Generally speaking, one only learns of a hidden barrier
after postulating its existence and testing it with a new calculation.
Detecting them is not straightforward and often involves a good deal
of CV space exploration.  Methods such as TAMD and well-tempered
metadynamics offer this capability, but much more work could be done
in the automated detection of hidden barriers and the ``right'' CV's
(e.g.,~\cite{das2006low,Perilla2012,ceriotti2011simplifying}).

An obvious way of reducing the likelihood of hidden barriers is to use
increase the dimensionality of CV space.  TAMD is  well-suited
to this because it is a gradient method, but standard metadynamics,
because it is a histogram-filling method, is not.  A recent variant of
metadynamics termed ``reconnaissance
metadynamics''~\cite{tribello2010self} does have the capability of
handling high-dimensional CV spaces.  In reconnaissance metadynamics,
bias potential kernels are deposited at the CV space points identified
as centers of clusters detected and measured by an on-the-fly
clusterization scheme.  These kernels are hyperspherically symmetric
but grow as cluster sizes grow and are able to push a system out of a
CV space basin to discover other basins.  As such, reconnaissance
metadynamics is an automated way of identifying free-energy minima in
high-dimensional CV spaces.  It has been applied the identification of
configurations of small clusters of
molecules~\cite{Tribello2011exploring} and identification of
protein-ligand binding poses~\cite{Soederhjelm2012locating}.

\subsubsection{Some Common and Emerging Types of CV's}\label{sec:cv_types}

There are very few ``best practices'' codified for choosing CV's for
any given system.  Most CV's are developed ad hoc based on the
processes that investigators would like to study, for instance,
center-of-mass distance between two molecules for studying
binding/unbinding, or torsion angles for studying conformational
changes, or number of contacts for studying order-disorder
transitions.  Cartesian coordinates of centers of mass of groups of
atoms are also often used as CV's, as are functions of these
coordinates.

The potential energy $V(\xb)$ is also an example of a 1-D CV, and
there have been several examples of using it in CV-based enhanced
sampling methods, such as umbrella
sampling~\cite{bartels1998probability}, metadynamics
\cite{micheletti2004reconstructing} well-tempered metadynamics
\cite{bonomi2010enhanced}.  In a recent work based on steered MD, it
has been shown that also relevant reductions of the potential energy
(e.g. the electrostatic interaction free-energy) can be used as
effective CV's~\cite{do2013rna}.  The basic rationale for enhanced
sampling of $V$ is that states with higher potential energy often
correspond to transition states, and one need make no assumptions
about precise physical mechanisms.  Key to its successful use as a CV,
as it is for any CV, is a proper accounting for its entropy; i.e., the
classical density-of-states.

Coarse-graining of particle positions onto Eulerian fields was used
early on in enhanced sampling~\cite{Roitberg1991}; here, the value of
the field at any Cartesian point is a CV, and the entire field
represents a very high-dimensional CV.  This idea has been put to use
recently in the ``indirect umbrella sampling'' method of Patel et
al.~\cite{Patel2011} for computing free energies of solvation, and
string method (Sec.~\ref{sec:string_method}) calculations of lipid
bilayer fusion~\cite{Mueller2012}.  In a similar vein, there have been
recent attempts at variables designed to count the recurrency of
groups of atoms positioned according to given templates, such as
$\alpha$-helices paired $\beta$-strands~in
proteins~\cite{pietrucci2009collective}.

We finally mention the possibility of building collective variables
based on set of frames which might be available from experimental data
or generated by means of previous MD simulations. Some of these
variables are based on the idea of computing the distances between the
present configuration and a set of precomputed snapshots. These
distances, here indicated with $d_i$, where $i$ is the index of the
snapshot, are then combined to obtain a coarse representation of the
present configuration, which is then used as a CV.  As an example, one
might combine the distances as
\begin{equation}
\label{eq:pathcv-s}
s=\frac{\sum_ie^{-\lambda d_i}i}{\sum_ie^{-\lambda d_i}}
\end{equation}
If the parameter $\lambda$ is properly chosen, this function returns a
continuous interpolation between the indexes of the snapshots which
are closer to the present conformation. If the snapshots are disposed
along a putative path connecting two experimental structures, this CV
can be used as a path CV to monitor and bias the progression along the
path \cite{branduardi2007b}.  A nice feature of path CVs is
that it is straighforward to also monitor the distance from the
putative path. The standard way to do it is by looking at the distance
from the closest reference snapshot, which can be approximately
computed with the following continuous function:
\begin{equation}
\label{eq:pathcv-z}
z=-\lambda^{-1}\log \sum_ie^{-\lambda d_i}
\end{equation}
\cfa{This approach, modified to use internal coordinates, was used
  recently by Zinovjev et al. to study the aqueous phase reaction of
  pyruvate to salycilate, and in the CO bond-breaking/proton transfer in PchB~\cite{Zinovjev2012acollective}.}

A generalization to multidimensional paths
(i.e. sheets) can be obtained by assigning a generic vector $v_i$ to
each of the precomputed snapshot and computing its average
\cite{spiwok2011metadynamics}:
\begin{equation}
s=\frac{\sum_ie^{-\lambda d_i}v_i}{\sum_ie^{-\lambda d_i}}
\end{equation}

\section{Tempering Approaches}\label{sec:tempering}

``Tempering'' refers to a class of methods based on increasing the
temperature of an MD system to overcome barriers.  Tempering relies on
the fact that according to the Arrhenius law the rate at which
activated (barrier-crossing) events happen is strongly dependent on
the temperature.  Thus, an annealing procedure where the system is
first heated and then cooled allows one to produce quickly samples
which are largely uncorrelated. The root of all these ideas indeed
lies in the simulated annealing procedure
\cite{kirkpatrick1983optimization}, a well-known method successfully
used in many optimization problems.

\subsection{Simulated tempering}

Simulated annealing is a form of Markov-chain Monte Carlo sampling
where the temperature is artificially modified during the simulation.
In particular, sampling is initially done at a temperature high enough
that the simulation can easily overcome high free-energy barriers.
Then, the temperature is decreased as the simulation proceeds, thus
smoothly bringing the simulation to a local energy minimum.  In
simulated annealing, a critical parameter is the cooling speed.
Indeed, the probability to reach the global minimum grows as this
speed is decreased.

The search for the global minimum can be interpreted in the same way
as sampling an energy landscape at zero temperature. One could thus
imagine to use simulated annealing to generate conformations at, e.g.,
room temperature by slowly cooling conformations starting at high
temperature. However, the resulting ensemble will strongly depend on
the cooling speed, thus possibly providing \cfa{a\ } biased result. A
better approach \cfa{consists of the\ } the so\cfa{-}called simulated
tempering methods \cite{marinari1992simulated}. Here, a discrete list
of temperatures $T_i$, with $i\in 1\dots N$ are chosen \emph{a
  priori}, typically spanning a range going from the physical
temperature of interest \cfa{to\ } a temperature which is high enough to
overcome all relevant free energy barriers.  (Note that we do not
have to stipulate a CV-space in which those barriers live.)  Then, the
index $i$, which indicates at which temperature the system should be
simulated, is evolved with time. Two kind of moves are possible: (a)
normal evolution of the system at fixed temperature, which can be done
with a usual Markov Chain Monte Carlo or molecular dynamics and (b)
change of the index $i$ at fixed atomic coordinates. It is easy to
show that the latter can be performed as a Monte Carlo step with
acceptance equal to
\begin{equation}
\alpha=\min\left(
1,\frac{Z_j}{Z_i}e^{-\frac{U(x)}{k_BT_j}+\frac{U(x)}{k_BT_i}}
\right)
\end{equation}
where $i$ and $j$ are the indexes corresponding to the present
temperature and the new one.  The weights $Z_i$ should be choosen so
as to sample equivalently all the value of $i$.  It must be noticed
that also within molecular dynamics simulations only the potential
energy usually appears in the acceptance. This is due to the fact that
the velocities are typically scaled by a factor
$\sqrt{\frac{T_j}{T_i}}$ upon acceptance.  This scaling leads to a
cancellation of the contribution to the acceptance coming from the
kinetic energy. Ultimately, this is related to the fact that the
ensemble of velocities is analytically known \emph{a priori}, such
that it is possible to adapt the velocities to the new temperature
instantaneously.

Estimating \cfa{these\ } weights $Z_i$ is nontrivial and typically requires a
preliminary step. Moreover, if this estimate is poor the system could
spend no time at the physical temperature, thus spoiling the result.
Iterative algorithms for adjusting these weights have been proposed
(see e.g. \cite{park2007choosing}). We also observe that since the
temperature sets the typical value of the potential energy, an effect
much similar to that of simulated tempering with adaptive weights can
be obtained by performing a metadynamics simulation using the
potential energy as a CV (Sec.~\ref{sec:cv_types}).

\subsection{Parallel tempering}
A smart way to alleviate the issue of finding the correct weights is
that of simulating several replicas at the same time
\cite{hansmann1997parallel,sugita1999replica}.  Rather that changing
the temperature of a single system, the defining move proposal in
parallel tempering consists of a coordinate swap between two $T$-replicas
with acceptance probability
\begin{equation}
\alpha=\min\left(
1,e^{\left(
\frac{1}{k_BT_j}-\frac{1}{k_BT_i}
\right)
\left[U(\xb_i)-U(\xb_j)\right]
}
\right)
\end{equation}
\gb{This method is the root of a class of techniques collectively
known as ``replica exchange'' methods, and the latter name is often
used as a synonimous of parallel tempering.}
Notably, within this framework it is not necessary to precompute a set
of weights. Indeed, the equal time spent by each replica at each
temperature is enforced by the constraint that only pairwise swaps are
allowed. Moreover, parallel tempering has an additional advantage:
since the replicas are weakly coupled and only interact when exchanges
are attempted, they can be simulated on different computers without
the need of a very fast interconnection (provided, of course, that a
single replica is small enough to run on a single node).

The calculation of the acceptance is very cheap as it is based on the
potential energy which is often computed alongside force evaluation.
Thus, one could in theory exploit also a large number of virtual,
rejected exchanges so as to enhance statistical
sampling~\cite{frenkel2004speed,coluzza2005virtual}.  Since efficiency
of parallel tempering simulation can deteriorate if the stride between
subsequent exchanges is too large
\cite{sindhikara2008exchange,bussi2009simple}, a typical recipe is to
choose this stride as small as possible, with the only limitation of
avoiding extra costs due to replica synchronization.  One can push
this idea further and implement asynchronous versions of parallel
tempering, where overhead related to exchanges is minimized
\cite{gallicchio2008asynchronous,bussi2009simple}.  One should be
however aware that, especially at high exchange rate, artifacts coming
from e.g.~the use of wrong thermostating schemes could spoil the
results \cite{rosta2009thermostat,sindhikara2010exchange}.

Parallel tempering is popular in simulations of protein conformational
sampling~\cite{Vreede2005,Zhang2012Mu}, protein
folding~\cite{sugita1999replica,Zhou2003,Garcia2003,Im2003,Mei2012,Berhanu2013}
and aggregation~\cite{Kokubo2004,Oshaben2012}, due at least in part to
the fact that one need not choose CV's to use it, and CV's for
describing these processes are not always straightforward to
determine.

\subsection{Generalized replica exchange}

The difference between the replicas is not restricted to be a change
in temperature.  Any control parameter can be changed, and even the
expression of the Hamiltonian can be modified
\cite{sugita2000replica}.  In the most general case every replica is
simulated at a different temperature (and or pressure) and a different
Hamiltonian, and the acceptance reads
\begin{equation}
\alpha=\min\left(
1,\frac{e^{-\left(\frac{U_i(x_j)}{k_BT_i}+\frac{U_j(x_i)}{k_BT_j} \right)}}{
e^{-\left(\frac{U_i(x_i)}{k_BT_i}+\frac{U_j(x_j)}{k_BT_j} \right)}}
\right)
\end{equation}

Several recipes for choosing the modified Hamiltonian have been
proposed in the literature
\cite{fukunishi2002hamiltonian,liu2005replica,affentranger2006novel,fajer2008replica,xu2008hamiltonian,zacharias2008combining,vreede2009reordering,itoh2010replica,meng2010constant,terakawa2011easy,wang2011replica,zhang2012folding,bussi2013hamiltonian}.
Among these, a notable idea is that of solute tempering
\cite{liu2005replica,wang2011replica} which is used for the simulation
of solvated biomolecules. Here, only the Hamiltonian of the solute is
modified. More precisely, one could notice that a scaling of the
Hamiltonian by a factor $\lambda$ is completely equivalent to a
scaling of the temperature by a factor $\lambda^{-1}$. Hamiltonian
scaling however can take advantage of the fact that the total energy
of the system is an extensive property. Thus, one can limit the
scaling to the portion of the system which is considered to be
interesting and which has the relevant bottlenecks. With solute
tempering, the solute energy is scaled whereas the solvent energy is
left unchanged. This is equivalent to keeping the solute at a high
effective temperature and the solvent at the physical temperature.
Since in the simulation of solvated molecules most of the atoms belong
to the solvent, this turns in a much smaller modification to the
explored ensemble when compared with parallel tempering. In spite of
this, the effect on the solute resemble much that of increasing the
physical temperature.

A sometime\cfa{s-}overlooked subtlety in solute tempering is the choice for
the treatment of solvent-solute interactions. Indeed, whereas
solute-solute interactions are scaled with a factor $\lambda<1$ and
solvent-solvent interactions are not scaled, any intermediate choice
(scaling factor between $\lambda$ and 1) could intuitively make sense
for solvent-solute coupling.  In the original formulation, the authors
used a factor $(1+\lambda)/2$ for the solute-solvent interaction. This
choice however was later shown to be suboptimal
\cite{huang2007replica,wang2011replica}, and refined to be
$\sqrt{\lambda}$.  This latter choice appears to be more physically
sound, since it allows one to just simulate the biased replicas with a
modified force-field. Indeed, if one scales the charges of the solute
by a factor $\sqrt{\lambda}$, electrostatic interactions are changed
by a factor $\lambda$ for solute-solute coupling and $\sqrt{\lambda}$
for solute-solvent coupling. The same is true for Lennard-Jones terms,
albeit in this case it depends on the specific combination rules used.
Notably, the same rules for scaling were used in a previous work
\cite{affentranger2006novel}.  As a final remark, we point out that
solute tempering can be also used in a serial manner \emph{a l\`a}
\cfa{simulated\ } tempering, in a simulated solute tempering scheme
\cite{denschlag2009simulated}.

\subsection{General comments}

In general, the advantage of these tempering methods over
straighforward sampling can be rationalized as follows. A simulation
is evolved so as to sample a modified ensemble by e.g. raising
temperature or artificially modifying the Hamiltonian.  The change in
the ensemble could be drastic, so that trying to extract canonical
averages by reweighting from such a simulation would be pointless.
For this reason, a ladder of intermediate ensembles is built,
interpolating between the physical one (i.e. room temperature,
physical Hamiltonian) and the modified one. Then, transitions between
consecutive steps in this ladder (or, in parallel schemes, coordinate
swaps) are performed using a Monte Carlo scheme.  Assuming that the
dynamics of the most modified ensemble is ergodic, independent samples
will be generated every time a new simulation reaches the highest step
of the ladder. Thus, efficiency of these methods is often based on the
evaluation of the round trip time required for a replica to traverse
the entire ladder.

Tempering methods are thus relying on the ergodicity of the most
modified ensemble. This assumption is not always correct. A very
simple example is parallel tempering used to accelerate the sampling
over an entropic barrier.  Since the height of an entropic barrier
grows with the temperature, in this conditions the barrier in the most
modified ensembles are unaffected \cite{zuckerman2006second}.
Moreover, since a lot of time is spent in sampling states in
non-physical situations (e.g. high temperature), the overall
computational efficiency could even be lower than that of
straightforward sampling. Real applications are often in an
intermediate situation, and usefulness of parallel tempering should be
evaluated case by case.

The number of intermediate steps in the ladder can be shown to grow
with the square root of the specific heat of the system in the case of
parallel tempering simulations.  No general relationship can be drawn
in the case of Hamiltonian replica exchange, but one can expect
approximately that the number of replicas should be proportional to
the square root of the number of degrees of freedom affected by the
modification of the Hamiltonian. Thus, Hamiltonian replica exchange
methods could be much more effective than simple parallel tempering as
they allow the effort to be focused and the number of replicas to be
minimized.

Parallel tempering has the advantage that all the replicas can be
analyzed to obtain meaningful results, e.g., to predict the melting
curve of a molecule. This procedure should be used with caution,
especially with empirically parametrized potentials, which are often
tuned to be realistic only at room temperature. On the other hand,
Hamiltonian replica exchange often relies on unphysically modified
ensembles which have no interest but for the fact that they increase
ergodicity.

As a final note, we observe that data obtained at different
temperature (or with modified Hamiltonians) could be combined to
enhance statistics at the physical temperature \cite{chodera2007use}.
However, the effectiveness of this data recycling is limited by the
fact that high temperature replicas visit very rarely low energy
conformations, thus decreasing the amount of additional information
that can be extracted.

\section{Combinations and Advanced Approaches}

\subsection{Combination of tempering methods and biased sampling}

The algorithms presented in Section~\ref{sec:tempering} and based on
tempering are typically considered to be simpler to apply when
compared with those discussed in Section~\ref{sec:cv} and based on
biasing the sampling of selected collective variables. Indeed, by
avoiding the problem of choosing collective variables which properly
describe the reaction path, most of the burden of setting up a
simulation is removed.  However, this comes at a price: considering
the computational cost, tempering methods are extremely expensive.
This cost is related to the fact that they are able to accelerate all
degrees of freedom to the same extent, without an \emph{a priori}
knowledge of the sampling bottlenecks. In this sense, Hamiltonian
replica exchange methods are in an intermediate situation, since they
are typically less expensive than parallel tempering but allow to
embed part of the knowledge of the system in the simulation set up.

Because of the conceptual difference between tempering methods and
CV-based methods, these approaches can be easily and efficiently
combined. As an example, the combination of metadynamics and parallel
tempering can be used to take advantage of the known bottlenecks with
biased collective variables at the same time accelerating the overall
sampling with parallel tempering \cite{bussi2006free}.  In that work,
the free energy landscape for the folding of a small hairpin was
computed by biasing a small number of selected CVs (gyration radius
and the number of hydrogen bonds). These CVs alone are not enough to
describe folding, as can be easily shown by performing a metadynamics
simulation using these CVs. However, the combination with parallel
tempering allowed acceleration of all the degrees of freedom blindly
and reversible folding of the hairpin.  This combined approach also
improves the results when compared with parallel tempering alone,
since it accelerates exploration of phase-space.  Moreover, since
parallel tempering samples the unbiased canonical distribution, it is
very difficult to use it to compute free-energy differences which are
larger than a few $k_BT$.  The metadynamics bias can be used to
disfavor, e.g., the folded state so as to better estimate the
free-energy difference between the folded and unfolded states.

It is also possible to combine metadynamics with the solute tempering
method so as to decrease the number of required replicas and the
computational cost \cite{camilloni2008exploring}. As an alternative to
solute tempering, metadynamics in the well-tempered ensemble can be
effectively used to enhance the acceptance in parallel tempering
simulations and to decrease the number of necessary replicas
\cite{bonomi2010enhanced}.  This combination of parallel tempering
with well-tempered ensemble can be pushed further and combined with
metadynamics on a few selected degrees of freedom
\cite{deighan2012efficient}.  As a final note, bias exchange molecular
dynamics \cite{piana2007bias} combines metadynamics and replica
echange in a completely different spirit: every replica is run using a
different CV, thus allowing many CVs to be tried at the same time.
This technique has been succesfully applied to several problems.  For
a recent review, we refer the reader to Ref.~\cite{baft+12cpc}.

\subsection{Some methods based on TAMD}
\subsubsection{String method in collective variables}\label{sec:string_method}

The string method is generally an approach to find pathways of minimal
energy connecting two points in phase space~\cite{e2002string}.  When
working in CV's, the string method is used to find minimal free-energy
paths (MFEP's)~\cite{Maragliano2006jcp}.  String method calculations
involve multiple replicas, each representing a point $\zb_s$ in
CV space at position $s$ along a discretized string connecting two
points of interest (reactant and product states, say).  The forces on
each replica's $\zb_s$ are computed and their $\zb_s$'s updated, as in
TAMD, with the addition of forces that act to keep the $\zb$'s
equidistant along the string (so-called reparameterization forces):
\begin{equation}\label{eq:smcv_zmotion}
\bar{\gamma}\dot{z}_{j}(s,t) = \displaystyle\sum_{k}\biggl[\tilde{M}_{jk}(\mathbf{x}(s,t))\kappa[\theta_{k}(\mathbf{x}(s,t))-z_{k}(s,t)]\biggr] + \eta_{z}(t) + \lambda(s,t)\displaystyle\frac{\partial{}z_{j}}{\partial{}s}
\end{equation}
Here, $\tilde{M}_{jk}$ is the metric tensor mapping distances on the
manifold of atomic coordinates to the manifold of CV space, $\eta$ is
thermal noise and
$\lambda(s,t)\displaystyle\frac{\partial{}z_{j}}{\partial{}s}$
represents the reparameterization force tangent to the string that is
sufficient to maintain equidistant images along the string.  String
method has been used to study activation of the insulin-receptor
kinase~\cite{Vashisth2012b}, docking of insulin to its
receptor~\cite{Vashisth2013}, myosin~\cite{Ovchinnikov2011},  In these
examples, the update of the string coordinates is done at a lower
frequency than the atomic variables in each image.

In contrast, in the on-the-fly variant of string method in CV's, the
friction on the $\zb_s$'s is set high enough to make the effective
averaging of the forces approach the true mean forces, and the $\zb$
updates occur in lockstep with the $\xb$ updates of the MD
system~\cite{maragliano2007onthefly}.  Just as in TAMD, the atomic
variables obey an equation of motion like Eq.~\ref{eq:tamd_atomic}
tethering them to the $\zb_s$.  Stober and Abrams recently
demonstrated an implementation of on-the-fly string method to study
the thermodynamics of the normal-to-amyloidogenic transition of
$\beta$2-microglobulin~\cite{Stober2012a}.  Unique in this approach
was the construction of a single composite MD system containing 27
individual $\beta$2 molecules restrained to points on 3 $\times$ 3
$\times$ 3 grid inside a single large solvent box.  \cfa{Zinovjev et al.
used \gb{a combination of the on-the-fly string method and of path-collective variables
(see Equations~\ref{eq:pathcv-s} and ~\ref{eq:pathcv-z})}
in a quantum-mechanics/molecular-mechanics
approach to study a methyltransferase
reaction~\cite{Zinovjev2013toward}.}

\subsubsection{On-the-fly free energy parameterization}\label{sec:otf_fep}

Because TAMD provides mean-force estimates as it is exploring
CV space, it stands to reason that those mean forces could be used to
compute a free energy.  In contrast, in the single-sweep
method~\cite{Maragliano2008}, the TAMD forces are only used in the
CV space exploration phase, not the free-energy calculation itself.
Recently, Abrams and Vanden-Eijnden proposed a method for using TAMD
directly to {\em parameterize} a free energy; that is, to determine
the best set of some parameters $\lambdab$ on which a free energy of
known functional form depends~\cite{Abrams2012}:
\begin{equation}
F(\zb) = F(\zb;\lambdab^*)
\end{equation}
The approach, termed ``on-the-fly free energy parameterization'', uses forces 
from a running TAMD simulation to progressively optimize $\lambdab$ using
a time-averaged gradient error:
\begin{equation}
  \label{eq:otfp}
  E(\lambdab) = \frac1{2t} \int_0^t \left|\nabla_z F\left[\zb(s), \lambdab(t)\right] + \kappa
  \left[\theta(\xb(s)) - \zb(s)\right]\right|^2 ds,
\end{equation}
If constructed so \cfa{that\ } $F$ is linear in $\lambdab =
(\lambda_1,\lambda_2,\dots,\lambda_M)$, minimization of $E$ can be
expressed as a simple linear algebra problem
\begin{equation}
\sum_jA_{ij}\lambda_j=b_i,\ \ \ i=1,\dots,M
\end{equation}
and the running TAMD simulation provides progressively better
estimates of $A$ and $b$ until the $\lambdab$ converge.  In the cited
work, it was shown that this method is an efficient way to derive
potentials of mean force between particles in coarse-grained molecular
simulations as basis-function expansions.  It is currently being
investigated as a means to parameterize free energies associated with
conformational changes of proteins.

Chen, Cuendet, and Tuckermann developed a very similar approach that
in addition to parameterizing a free energy using d-AFED-computed
gradients uses a metadynamics-like bias on the
potential~\cite{Chen2012Tuckermann}.  These authors demonstrated
efficient reconstruction of the four-dimensional free-energy of vacuum
alanine dipeptide with this approach.

\section{Concluding Remarks}

In this review, we have summarized some of the current and emerging
enhanced sampling methods that sit atop MD simulation.  These have
been broadly classified as methods that use collective variable
biasing and methods that use tempering.  CV biasing is a much more
prevalent approach than tempering, due partially to the fact that it
is perceived to be cheaper, since tempering simulations are really
only useful for enhanced sampling of configuration space when run in
parallel.  CV-biasing also reflects the desire to rein in the
complexity of all-atom simulations by projecting configurations into a
much lower dimensional space.  (Parallel tempering can be thought of
as increasing the dimensionality of the system by a factor equal to
the number of simulated replicas.)  But the drawback of all CV-biasing
approaches is the risk that the chosen CV space does not provide the
most faithful representation of the true spectrum of metastable
subensembles and the barriers that separate them.  Guaranteeing that
sampling of CV space is not stymied by hidden barriers must be of
paramount concern in the continued evolution of such methods.  For
this reason, methods that specifically allow broad exploration of
CV space, like TAMD (which can handle large numbers of CV's) and
well-tempered metadynamics will continue to be valuable.  So too will
parallel tempering because its broad sampling of configuration space
can be used to inform the choice of better CV's.  Accelerating
development of combined CV-tempering methods bodes well for enhanced
sampling generally.

Although some of these methods involve time-varying forces (ABF, TAMD,
and metadynamics), all methods we've discussed have the underlying
rationale of the equilibrium ensemble.  TI uses the constrained
ensemble, ABF and metadynamics ideally converge to an ensemble in
which a bias erases free-energy variations, and TAMD samples an
attenuated/mollified equilibrium ensemble.  There is an entirely
separate class of methods that inherently rely on {\em
  non-equilibrium} thermodynamics.  We have not discussed at all the
several free-energy methods based on non-equilibrium MD simulations;
we refer interested readers to the article by Christoph Dellago in
this issue.

Finally, we have also not really touched on any of the practical
issues of implementing and using these methods in conjunction with
modern MD packages (e.g., NAMD~\cite{Phillips2005},
LAMMPS~\cite{Plimpton1995}, Gromacs~\cite{Hess2005gromacs},
Amber~\cite{Case2005amber}, and CHARMM~\cite{Brooks1983charmm}, to
name a few).  At least two packages (NAMD and CHARMM) have native
support for collective variable biasing, and NAMD in particular offers
both native ABF and a TcL-based interface which has been used to
implement TAMD~\cite{Abrams2010}.  The native collective variable
module for NAMD has been recently ported to
LAMMPS~\cite{fiorin2013using}.  Gromacs offers native support for
parallel tempering.  Generally speaking, however, modifying MD codes
to handle CV-biasing and multiple replicas is not straightforward,
since one would like access to the data structures that store
coordinates and forces.  A major help in this regard is the PLUMED
package~\cite{bonomi2009,tribello2013plumed2}, which patches a
variety of MD codes to enable users to use many of the techniques
discussed here.

\cfa{
\section{Abbreviations}
\begin{itemize}
\item
{\bf ABF}: adaptive-biasing force
\item
{\bf AFED}: adiabatic free-energy dynamics
\item
{\bf CV}: collective-variable
\item
{\bf MD}:  molecular dynamics
\item
{\bf MFEP}:  minimum free-energy path
\item
{\bf TAMD}: temperature-accelerated molecular dynamics
\item
{\bf TI}: thermodynamic integration
\item
{\bf WHAM}: weighted-histogram analysis method
\end{itemize}
}

\section*{Acknowledgments}

CFA would like to acknowledge support of NSF (DMR-1207389) and NIH (1R01GM100472). 
GB would like to acknowledge the European Research Council (Starting Grant S-RNA-S, no.~306662)
for financial support.  Both authors would like to acknowledge NSF support of a recent Pan-American Advanced Studies Institute Workshop ``Molecular-based Multiscale Modeling and Simulation'' (OISE-1124480; PI: W. J. Pfaednter, U. Washington) held in Montevideo, Uruguay, Sept. 12-15, 2012, where the authors met and began discussions that influenced the content of this review. 

\bibliographystyle{mdpi}
\makeatletter
\renewcommand\@biblabel[1]{#1. }
\makeatother

\end{document}